\documentclass[twocolumn,superscriptaddress,nofootinbib,amsmath,amssymb,prc]{revtex4-2}

\usepackage{xcolor}
\usepackage{graphicx}
\usepackage{dcolumn}
\usepackage{bm}
\usepackage{ifpdf}
\usepackage{epstopdf}
\usepackage{slashed}
\usepackage{amsfonts}
\usepackage{mathrsfs}
\usepackage{amssymb}
\usepackage{multirow}
\usepackage{CJK}
\usepackage{xcolor}
\usepackage{subfigure,dcolumn}
\usepackage{times}
\usepackage[T2A,T1]{fontenc}
\usepackage[russian,english]{babel}
\usepackage{amsmath}
\usepackage{listings}
\usepackage{booktabs}
\usepackage{color}
\usepackage[misc]{ifsym}
\usepackage{float}
\usepackage{hyperref}

\usepackage{physics}

\makeatletter
\def\NAT@def@citea{\def\@citea{\NAT@separator}}
\makeatother

\begin{document}


\title{Time-dependent random phase approximation for particle-number fluctuations and correlations\\
in deep-inelastic collisions of $^{144}$Sm+$^{144}$Sm and $^{154}$Sm+$^{154}$Sm}

\author{Zepeng Gao}
\affiliation{Sino-French Institute of Nuclear Engineering and Technology, Sun Yat-sen University, Zhuhai 519082, China}
\affiliation{Department of Physics, School of Science, Institute of Science Tokyo, Tokyo 152-8551, Japan}

 \author{Kazuyuki Sekizawa}
 \email[Corresponding author, ]{sekizawa@phys.sci.isct.ac.jp}
\affiliation{Department of Physics, School of Science, Institute of Science Tokyo, Tokyo 152-8551, Japan}
\affiliation{Nuclear Physics Division, Center for Computational Sciences, University of Tsukuba, Ibaraki 305-8577, Japan}
\affiliation{RIKEN Nishina Center, Saitama 351-0198, Japan}

\author{Long Zhu}
\email[Corresponding author, ]{zhulong@mail.sysu.edu.cn}
\affiliation{Sino-French Institute of Nuclear Engineering and Technology, Sun Yat-sen University, Zhuhai 519082, China}
\affiliation{Guangxi Key Laboratory of Nuclear Physics and Nuclear Technology, Guangxi Normal University, Guilin 541004, China}

\date{\today}

\begin{abstract}
\edef\oldrightskip{\the\rightskip}
\begin{description}
\rightskip\oldrightskip\relax
\setlength{\parskip}{0pt} 
\item[Background]
The fluctuation-dissipation mechanism underlying non-equilibrium transport in low-energy heavy-ion reactions remains unclear. Although the time-dependent Hartree-Fock (TDHF) method provides a reasonable description of average reaction outcomes and one-body dissipation, it is known to significantly underestimate fluctuations of observables.

\item[Purpose]
The purpose of this work is to investigate deep-inelastic collisions of $^{144}$Sm+$^{144}$Sm and $^{154}$Sm+$^{154}$Sm with microscopic mean-field approaches and to show a predominant role of one-body dissipation as well as one-body fluctuations and correlation in low-energy heavy-ion reactions.

\item[Methods]
Three dimensional TDHF calculations are carried out for $^{144}$Sm+$^{144}$Sm at $E_\text{c.m.}=500$\,MeV and $^{154}$Sm+$^{154}$Sm at $E_\text{c.m.}=485$\,MeV for a range of impact parameters with Skyrme SLy5 energy density functional. Backward time evolutions are performed as well to evaluate fluctuations and correlation in nucleon numbers within time-dependent random phase approximation (TDRPA).
 
\item[Results]
As results of TDHF calculations we obtain total kinetic energy loss (TKEL), scattering angles, and contact time, for a wide range of impact parameters. TKEL takes almost constant values in an impact parameter range of $0\le b\lesssim6$\,fm, indicating a fully-damped character of the reactions. We find a systematically lower TKEL for $^{144}$Sm+$^{144}$Sm collisions as compared with the other, presumably because of the shell effects of $N=82$. With TDRPA we calculate mass- and charge-number fluctuations, $\sigma_{AA}$ and $\sigma_{ZZ}$, as well as the correlation between neutron and proton transfers, $\sigma_{NZ}$, for each impact parameter. By combining these results, we make a comparison of the $\sigma^2$-TKEL plot with available experimental data. We demonstrate that TDRPA quantitatively reproduces the experimental $\sigma_{AA}^2$-TKEL distributions, whereas it systematically underestimates the charge fluctuation, $\sigma_{ZZ}$. The double-differential cross sections of reaction products are calculated, showing good agreement with the experimental data. We confirm a long-thought characteristic property that the closed-shell structure limits nucleon transfer at small energy losses, based on our microscopic TDHF and TDRPA calculations.

\item[Conclusions]
It has been shown that one-body fluctuations and correlation in TDRPA, built on top of TDHF mean-field dynamics, are the predominant mechanism in deep-inelastic collisions. TDRPA systematically overestimates proton-number fluctuations $\sigma_{ZZ}$ in both $^{144}$Sm+$^{144}$Sm and $^{154}$Sm+$^{154}$Sm reactions, which offers a room for refining our understanding of reaction mechanisms in low-energy heavy-ion reactions.

\end{description}
\end{abstract}
\maketitle

\section{Introduction}
\label{sec:Intro}

In deep-inelastic collisions, a substantial portion of the relative kinetic energy between the collision partners is dissipated into intrinsic degrees of freedom, typically accompanied by extensive nucleon transfer \cite{freiesleben1984nz}. It is generally assumed that the dominant mechanism for energy dissipation is one-body dissipation, wherein energy is transferred through the sequential exchange of independent nucleons, with each nucleon imparting recoil momentum \cite{huizenga1976energy}. The nucleon exchange induces dispersion (denoted as $\sigma^2$, where $\sigma$ corresponds to fluctuation) in the particle numbers of the fragments, and $\sigma^2$ become increasingly pronounced with greater total kinetic energy loss (TKEL). However, at small TKEL (grazing collisions), many experimental data deviate from the systematic $\sigma^2$–TKEL relationship predicted solely by sequential exchange of independent particles \cite{dakowski1980structure,schroder1977dissipation,liao2024shell}. This deviation is clearly influenced by the strong driving potential of the potential energy surface of the dinuclear system in the $N$-$Z$ plane, including shell effects, that leads to fast isospin equilibration \cite{wu1981influence,hildenbrand1983influence,gobbi1981evolution}. Moreover, the average free path of nucleons is limited by the Pauli blocking during dissipation processes, which in turn influences particle-number fluctuations \cite{randrup1978mass,wilcke1980bombarding,schroder1980effect}. Overall, the fluctuation-dissipation mechanism underlying non-equilibrium transport in many-body quantum systems remains unclear. 

On the other hand, multinucleon transfer (MNT) reactions, which often accompany substantial energy dissipation and could also be classified as deep-inelastic collisions, show promise for synthesizing new isotopes far from the $\beta$-stability line, and may even open promising avenues for reaching the ``island of stability'' for superheavy elements \cite{volkov1978deep,zagrebaev2008production,sekizawa2017enhanced,ZHU2022137113}. One of the primary challenges in synthesizing new isotopes via MNT reactions is minimizing energy dissipation during the transfer of large numbers of nucleons, thereby enhancing the survival probability of the reaction products \cite{zagrebaev2013production,YANG2025139318}. Addressing this challenge also requires a deeper understanding of the interplay between nucleon transfer and energy dissipation. This characteristic should also be taken into consideration in the adopted entrance channel conditions. For instance, the U+U system \cite{hildenbrand1977reaction,freiesleben1979reaction}, under identical energy loss conditions, exhibits a markedly higher $\sigma^2_{ZZ}$ compared to the Pb+Pb system \cite{tanabe1980pb}, suggesting that the average energy loss per exchanged proton is lower. These differences can likely be attributed to the shell effects of both the target and projectile nuclei, as well as the average binding energy of the valence nucleons. 

In recent years, a substantial body of research on MNT reactions has emerged based on both macroscopic and microscopic models \cite{PhysRevC.107.014614,LI2020135697,PhysRevC.109.054612,PhysRevC.87.034608,PhysRevC.103.024608,PhysRevC.107.024604,PhysRevC.109.044617,PhysRevC.99.014613,yangyu,PhysRevC.111.024605,PhysRevResearch.5.L022021,PhysRevC.106.L011602,PhysRevC.111.024603,PhysRevC.107.054613,PhysRevC.110.044602,PhysRevC.110.054607,PhysRevC.111.014603,ZHU2024138423,ZHU2021136226,PhysRevC.110.064609,PhysRevC.110.054601,PhysRevC.108.024602,PhysRevC.106.054601}. Although these models can adequately reproduce many experimental observations, they generally fail to capture the more subtle structural effects. Time-dependent mean-field approaches, such as time-dependent Hartree-Fock (TDHF) or time-dependent density functional theory (TDDFT), have shown remarkable successes in describing nuclear excitations and dynamics microscopically \cite{RevModPhys.54.913,simenel2012nuclear,SIMENEL201819,stevenson2019low,sekizawa2019tdhf}. Especially, recent TDHF calculations (with or without addition of pairing correlations) have shown that the main (or average) reaction outcomes can be described quantitatively without adjustable parameters, while self-consistently incorporating both static and dynamical structural effects. Further developments of the theoretical framework and its applications can thus be considered to be promising to develop our understanding of complex reaction mechanisms in low-energy heavy-ion reactions. 

However, there is a well-known, longstanding drawback inherent in the standard TDHF approach, that is, it severely underestimates the width of fragment mass and charge distributions \cite{RevModPhys.54.913}. Recent theoretical efforts have shown that this drawback can be overcome by incorporating one-body fluctuations and correlations on top of the average (TDHF) trajectory, based on, \textit{e.g.}, time-dependent random phase approximation (TDRPA) \cite{simenel2012nuclear} or stochastic mean-field theory (SMF) \cite{AYIK2008174,lacroix2014stochastic}. Although those approaches have shown successful reproductions of existing experimental data \cite{PhysRevLett.54.1139,PhysRevLett.106.112502,PhysRevLett.120.022501,PhysRevC.102.014620,PhysRevC.102.064619,PhysRevC.97.054618,PhysRevC.100.014609,PhysRevC.100.044614,PhysRevC.104.054614,PhysRevC.107.014609,PhysRevC.108.054605,PhysRevC.108.064604}, we need further systematic calculations in comparison with available experimental data, to unveil underlying reaction mechanisms.

To this end, we have conducted TDRPA calculations for $^{144}$Sm+$^{144}$Sm (spherical\,+\,spherical) and $^{154}$Sm+$^{154}$Sm (deformed\,+\,deformed) systems for which old, yet great experimental data are available \cite{wu1981influence,hildenbrand1983influence}. The choice of symmetric systems is particularly important for conducting TDRPA calculations, because it has been recognized that TDRPA, at least in the present formulation, cannot be applied for asymmetric systems~\cite{PhysRevLett.120.022501}. The experimental data include TKE-$A$, $\sigma^2$-TKEL, as well as mass and charge distributions, allowing us to make a detailed comparison between the theory and experiments. This paper provides essential information on applicability of the TDRPA approach which incorporates one-body dissipation, fluctuations and correlations, on top of the TDHF mean-field dynamics.

The article is organized as follows. In Sec.~\ref{sec:method}, we will introduce theoretical framework of the TDHF and TDRPA approaches, and computational settings are presented in Sec.~\ref{sec:setting}. The correlation including the structural effects between particle-number fluctuations and energy dissipation, as well as the double-differential cross sections of reaction products are discussed in detail in Sec.~\ref{sec:result}. The conclusions and outlook are presented in Sec.~\ref{sec:conclusion}.


\section{METHODS: TDHF AND TDRPA}
\label{sec:method}

The TDHF is a microscopic dynamical approach that has been extensively employed in the study of low-energy nuclear reactions \cite{RevModPhys.54.913,simenel2012nuclear,SIMENEL201819,stevenson2019low,sekizawa2019tdhf}. We begin with the variation of the action expressed as:
\begin{equation}
\label{eq:1}
S = \int_{t_1}^{t_2} \langle \Psi(t) | (i\hbar\dv{t} - \hat{H}) | \Psi(t) \rangle \, \dd t, 
\end{equation}
where \( S \) is the action, \( |\Psi(t)\rangle \) is the many-body wave function, and \(\hat{H}\) is the Hamiltonian. The wave function \( |\Psi(t)\rangle \) in the TDHF framework is typically represented as a single Slater determinant composed of single-particle orbitals:
\begin{equation}
\label{eq:2}
|\Psi(t)\rangle = \frac{1}{\sqrt{N!}} \det\{ \psi_i(\boldsymbol{r}\sigma q,t)\},
\end{equation}
where \(A\) is the number of nucleons, and $\psi_i(\boldsymbol{r}\sigma q,t)$ denote the time-dependent single-particle wave functions with spatial coordinate $\boldsymbol{r}$, spin $\sigma$, and isospin $q$, at time $t$. 
By performing the variation of the action with respect to the single-particle wave functions $\psi_i^*(\boldsymbol{r}\sigma q,t)$, one can derive the equations of motion for the single-particle states. This variation leads to a set of coupled equations, which are essentially the time-dependent Schr\"odinger equations for each single-particle state in the self-consistent mean field:
\begin{equation}
\label{eq:3}
i\hbar\frac{\rm d}{{\rm d}t} \psi_i(\boldsymbol{r}\sigma q,t) = \hat{h}(t) \psi_i(\boldsymbol{r}\sigma q,t),
\end{equation}
where \( \hat{h}(t) \) is the single-particle Hamiltonian that includes the mean field generated by all the other nucleons. These nonlinear equations are solved using a three-dimensional Cartesian grid, enabling a thorough analysis without imposing any symmetry restrictions.

As mentioned earlier, TDHF can accurately capture the average behavior, but severely underestimates fluctuations \cite{RevModPhys.54.913}. This is primarily due to the squared operators, required to compute fluctuations of one-body observable expressed by $\hat{X}$ as 
\begin{equation}
\label{eq:4}
\sigma_{XX}=\sqrt{\langle\hat{X}^2\rangle-\langle\hat{X}\rangle^2},
\end{equation}
which lies outside the variational space of the TDHF framework \cite{PhysRevLett.47.1353}. Moreover, under the constraints of the least-action principle in TDHF, the collective trajectory of the system remains a uniquely classical path, even though the single-particle wave function embodies quantum characteristics. As a result, the TDHF wave function after collision becomes a superposition of states with different transfer channels. With the particle-number projection (PNP) method \cite{simenel2010particle}, one can indeed extract transfer probabilities and it has been extensively applied in the calculation of production cross sections in MNT reactions \cite{sekizawa2013time,sekizawa2014particle,sekizawa2016time,sekizawa2017microscopic,jiang2020probing,PhysRevC.109.024614,WU2022136886,PhysRevC.100.014612,gao2024role}. However, the PNP method exhibits varying degrees of inadequacy when describing the net transfer of large numbers of nucleons or capturing neutron-proton correlations \cite{PhysRevC.102.014620}. This limitation is caused by the use of the single Slater determinant that hinders the proper amplification of fluctuations during TDHF evolution, ultimately leading to an underestimation of the distribution width of observables \cite{reinhard1980time,broomfield2008mass}.

In fact, Balian and Vénéroni proposed in 1981 an action-like quantity expressed as:  
\begin{equation}
\label{eq:5}
\begin{aligned}
J & =\mathrm{Tr}\left[\hat{A}(t_1)\hat{D}(t_1)\right] \\
 & \quad-\int_{t_0}^{t_1}\mathrm{Tr}\left[\hat{A}(t)\left(\dv{\hat{D}(t)}{t}+\frac{i}{\hbar}\bigl[\hat{H}(t),\hat{D}(t)\bigr]\right)\right]\dd t,
\end{aligned}
\end{equation}
in order to incorporate the observable of interest into the variational space \cite{PhysRevLett.47.1353,BALIAN1984301,BALIAN1992351}. $\hat{D}(t)$ is the density matrix representing the state of the system, and $\hat{A}(t)$ is the observable, which respectively satisfy the boundary conditions:
\begin{equation}
\label{eq:6}
\hat{D}(t_0)=\hat{D}_0,\;\;\hat{A}(t_1)=\hat{A}_1.
\end{equation}
$t_0$ and $t_1$ represent the initial and final times of the evolution, respectively. With the boundary conditions, performing $\delta_DJ=0$ leads to the evolution equation of obsevable as:
\begin{equation}
\label{eq:7}
i\hbar\dv{\hat{A}(t)}{t} = \bigl[\hat{H}, \hat{A}(t)\bigr],
\end{equation}
which is the time reversal of the Heisenberg equation of motion. Under this variational space, the fluctuations/correlations of the observable can be represented as:
\begin{equation}
\label{eq:8}
\begin{aligned}
C_{ij}(t_1)
=& \lim_{\varepsilon_i, \varepsilon_j \to 0}
\frac{1}{2\,\varepsilon_i\,\varepsilon_j}
\,\mathrm{Tr}\Bigl[
\bigl(\rho^{(0)}(t_0) - \eta_i(t_0,\varepsilon_i)\bigr)\\
&\bigl(\rho^{(0)}(t_0) - \eta_j(t_0,\varepsilon_j)\bigr)
\Bigr],
\end{aligned}
\end{equation}
where the single particle matrices $\eta(t,\varepsilon)$ obey the TDHF equation with a boundary condition defined at final time $t_1$ as:
\begin{equation}
\label{eq:9}
\eta_j(t_1, \varepsilon_j)
= e^{i\,\varepsilon_j\,X_j}\,\rho^{(0)}(t_1)\,e^{-\,i\,\varepsilon_j\,X_j}.
\end{equation}
Here, $X_j$ is the one-body operator extracted from $\hat{A}\equiv e^{-\sum_j\varepsilon_j\hat{X}_j}$. $\varepsilon$ permits the incorporation of possible fluctuations in the small amplitude limit (RPA-level) within the TDHF mean-field evolution. The solution of the variational principle with this choice of variational spaces (\ref{eq:5}) and boundary
conditions given in Eq.~(\ref{eq:6}) could be found in Refs.~\cite{BALIAN1992351,simenel2012nuclear,broomfield2009calculations}. In Eq.~(\ref{eq:9}), the phase multiplication generates fluctuations that are propagated backward from \(t_1\) to \(t_0\) in the Heisenberg picture. Consequently, to calculate the correlations, the state at \(t_1\) is evolved backward to the initial time \(t_0\), which explains why the correlations at \(t_1\) depend on the density matrices defined at \(t_0\). Futhermore, if the backward propagated trajectories share the same mean field as the forward evolution, the TDRPA fluctuations reduce to the intrinsic TDHF fluctuations; otherwise, $C_{ij}$ will be amplified over time.

In this work, the fluctuations and correlation in nucleon transfers in deep-inelastic collisions are mainly investigated, where one can use the particle-number operator for a subspace $V$,
\begin{equation}
\label{eq:10}
\hat{X}=\hat{N}^{(q')}_V 
=\sum_{i,\sigma,q=q'} \int \hat{a}_i^\dagger(\mathbf{r}\sigma q)\hat{a}_i(\mathbf{r}\sigma q)\Theta(\mathbf{r}) \dd\mathbf{r},
\end{equation}
where $\Theta(\mathbf{r})$ = 1 in the volume $V$ that contains one of reaction products, and 0 elsewhere. In accordance with Eq.~(\ref{eq:9}), at time \(t_1\) one can transform the single-particle orbitals as:
\begin{equation}
\label{eq:11}
\phi_i^{(X)}\bigl(\mathbf{r}\sigma q, t_1)=e^{-i\varepsilon\hat{X}}
\psi_{i}\bigl(\mathbf{r}\sigma q, t_1\bigr).
\end{equation}
Hence, the fluctuations/correlations in Eq.~(\ref{eq:8}) can be rewritten as:
\begin{equation}
\label{eq:12}
\sigma_{XY}(t_1)= \sqrt{\frac{\eta_{00}(t_0)+\eta_{XY}(t_0)-\eta_{0X}(t_0)-\eta_{0Y}(t_0)}{2\,\varepsilon^2}},
\end{equation}
where we set $\varepsilon_i=\varepsilon_j=\varepsilon$ and
\begin{eqnarray}
&&\eta_{XX'}=\sum_{i,j=1}^A\sum_{\sigma}\left|\left\langle \phi_{i}^{(X)}(\mathbf{r}\sigma q,t_0)\middle|\phi_{j}^{(X')}(\mathbf{r}\sigma q,t_0) \right\rangle\right|^2\label{eq:13}
\end{eqnarray}
with
\begin{eqnarray}
&&\left\langle \phi_{i}^{(X)}(\mathbf{r}\sigma q,t_0)\middle|\phi_{j}^{(X')}(\mathbf{r}\sigma q,t_0) \right\rangle \nonumber\\[1.5mm]
&&\equiv \sum_\sigma\int \phi_i^{(X)*}(\boldsymbol{r}\sigma q,t_0)\phi_j^{(X')}(\boldsymbol{r}\sigma q,t_0)\dd\boldsymbol{r}.
\end{eqnarray}
The subscript `0' in Eq.~(\ref{eq:12}) means that $\phi_{i}^{(0)}(\mathbf{r}\sigma q,t_0)$ is computed by the backward evolution using $\hat{X}=0$. By changing $X$ and $Y$ in Eq.~\eqref{eq:12}, one can obtain neutron- and proton-number fluctuations, $\sigma_{NN}$ and $\sigma_{ZZ}$, as well as the correlation $\sigma_{NZ}$. The mass fluctuation of a fragment can be calculated as
\begin{equation}
\label{eq:14}
\sigma_{AA}^2=\sigma_{NN}^2+\sigma_{ZZ}^2+2\sigma^2_{NZ}.
\end{equation}

\section{COMPUTATIONAL DETAILS}
\label{sec:setting}

In this work, the calculation of static nuclei and the dynamic reaction process are described coherently by static HF and TDHF methods using the Sky3D code \cite{schuetrumpf2018tdhf}. The code has been extended to carry out the particle-number projection \cite{gao2024role} as well as TDRPA. The Skyrme energy density functional SLy5 \cite{chabanat1998skyrme} has been utilized in both static and dynamic processes, as well as TDRPA calculations. The static HF calculations are performed using the damped gradient iteration method, and the box grid points are established to be 30 × 30 × 30 with a mesh spacing of 0.8 fm, while 70 × 30 × 70 of box grid points for dynamical simulation are further fixed. The time step is set to be $\Delta t$ = 0.2 fm/$c$.

The Hartree-Fock ground state of $^{154}_{\phantom{1}62}$Sm$_{92}^{}$ is deformed in a prolate shape with $\beta_2 \approx 0.32$ and with a small $\beta_3$, while that of $^{144}_{\phantom{1}62}$Sm$_{82}^{}$ is of spherical shape, because of the $N=82$ shell closure. Note that the latest experiment has precisely measured the deformation parameter $\beta_2 = 0.2925(25)$ for $^{154}$Sm \cite{PhysRevLett.134.022503}. In the dynamic calculations, the projectile and target nuclei are placed in the computational box boosted with proper momentum on the Rutherford trajectory. We take the collision axis parallel to the $x$ axis, while the impact parameter vector is set along positive $y$ direction, thus the reaction plane is the $xy$ plane. For the $^{154}$Sm+$^{154}$Sm (defomed\,+\,deformed) system, three orientations, \textit{i.e.} Tip-Tip, Side-Side, and Tip-Side, were considered. In the Tip-Tip (Side-Side) configuration, the symmetry axis of two $^{154}$Sm nuclei are set parallel to $x$ ($y$) axis at the initial time. In the Tip-Side configuration, the symmetry axes of projectile and target nuclei are set parallel to $x$ and $y$ axes, respectively. The center-of-mass energies were set at 500 MeV for $^{144}$Sm+$^{144}$Sm and 485 MeV for $^{154}$Sm+$^{154}$Sm to facilitate comparison with the existing experimental data \cite{wu1981influence,hildenbrand1983influence}. For each system, 50 sets of TDHF time evolutions were simulated with an impact parameter increment of 0.2 fm. For each TDHF time evolution, three backward time evolutions were performed: one for neutron, one for proton, and one without any modification for $\psi_i(\mathbf{r}\sigma q, t_1)$ \eqref{eq:11}. Consequently, this work simulated a total of 800 time evolutions, with an additional 150 simulations conducted to investigate the dependence on $\varepsilon$. The choices of $\varepsilon = 10^{-2}$, $10^{-3}$, and $10^{-4}$ had little impact on the results. In the following, we show results obtained with $\varepsilon = 10^{-3}$.

\section{COMPUTATIONAL RESULTS}
\label{sec:result}

\subsection{Average outcomes of TDHF}

\begin{figure}[t]
\includegraphics[width=0.46\textwidth]{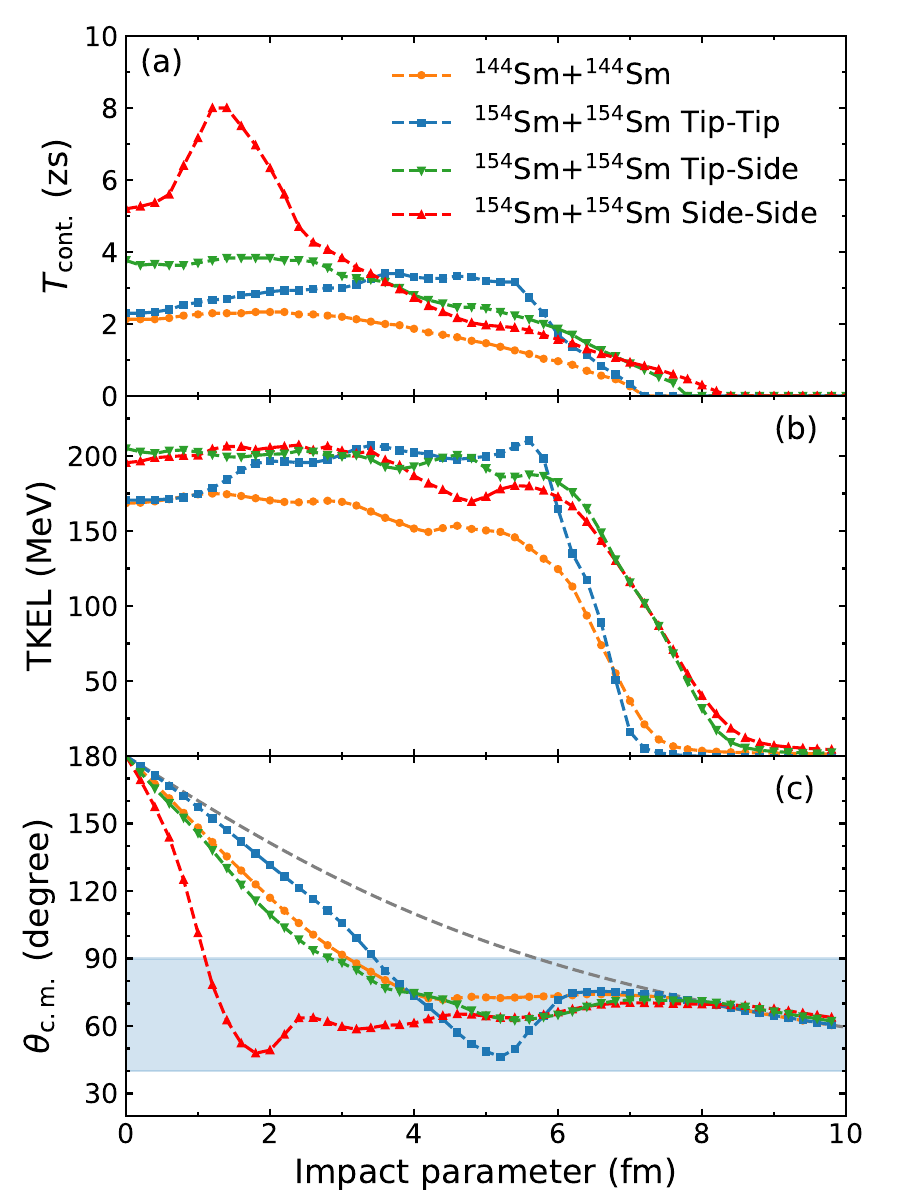}
\caption{\label{fig:1} Contact time (a), total kinetic energy loss (TKEL) (b), and scattering angle in the center-of-mass frame (c) are shown as functions of the impact parameter. Results for $^{144}$Sm+$^{144}$Sm, $^{154}$Sm+$^{154}$Sm with orientation of Tip-Tip, Side-Side, and Tip-Side are shown by orange circles, blue squares, red upper triangles, green lower triangles, respectively. In the panel (c), the scattering angle for the Rutherford trajectory is also shown by gray dashed curve and the shaded area represents the angular coverage of the experiments \cite{wu1981influence,hildenbrand1983influence}.}
\end{figure}

First, let us examine average reaction dynamics obtained from TDHF caluclations. Figure~\ref{fig:1} presents the impact parameter dependence of three critical observables: total contact time, TKEL, and scattering angle. The contact time is defined as the duration during which the minimum density between the colliding nuclei exceeds $\rho = 0.08$\,fm$^{-3}$. Our analysis reveals distinct dynamics depending on the initial orientations: Side–Side collisions at small impact parameters ($b \lesssim 4$\,fm) produce compact dinuclear configurations, while Tip–Tip collisions develop an elongated dinuclear system and it subsequently ruptures. This structural effect naturally explains the significantly prolonged contact times observed in Side–Side collisions compared to the Tip–Tip case at small $b$-values. Correspondingly, Side–Side configurations predominantly yield forward-peaked angular distributions at small $b$-values. However, the system exhibits remarkable behavioral inversion at larger impact parameters ($b \gtrsim 4$ fm) that the underlying reaction mechanisms undergo a transition, leading to an interchange of the observed behaviors between the two configurations \cite{PhysRevC.93.054616}. Intermediate configurations (Tip–Side) and the $^{144}$Sm+$^{144}$Sm system display transitional characteristics that interpolate between these extremes. While TKEL demonstrates limited sensitivity to interaction duration in small impact parameter region, a pronounced correlation emerges in grazing collisions where reduced reaction times correspond to systematically lower TKEL. These orientation-dependent effects align basically with prior theoretical investigations \cite{PhysRevC.93.054616}, which is consistent with our conclusions. The blue-shaded region in Fig.~\ref{fig:1}(c) indicates the angular range covered by experimental detection systems \cite{wu1981influence,hildenbrand1983influence}, and subsequent quantitative comparisons with theoretical values will incorporate impact parameter cut constrained by this angular acceptance.

\begin{figure}[t]
\includegraphics[width=0.48\textwidth]{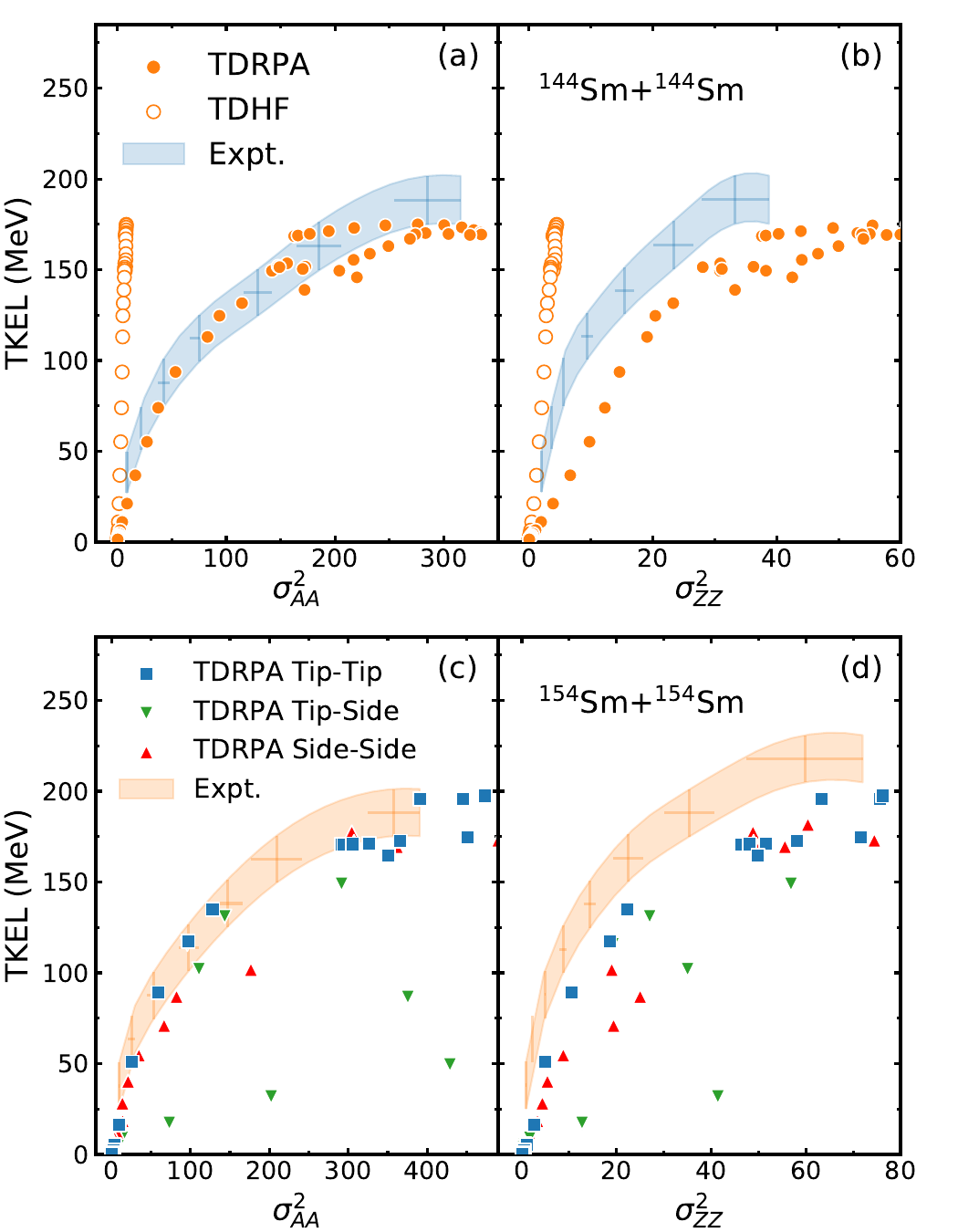}
\caption{\label{fig:2} In panels (a) and (b), TKEL is plotted as a function of mass and charge dispersions, $\sigma^2_{AA}$ and $\sigma^2_{ZZ}$, respectively, for collisions of $^{144}$Sm+$^{144}$Sm at $E_{\rm c.m.} = 500$\,MeV. TDRPA and TDHF results are shown by solid and open circles, respectively. Panels (c) and (d) are the same as upper panels, but for collisions of $^{154}$Sm+$^{154}$Sm at $E_{\rm c.m.} = 485$\,MeV. TDRPA results for the Tip-Tip, Side-Side , and Tip-Side orientations are displayed as blue squares, red upper triangles, and green lower triangles, respectively. Experimental data are shown by solid lines with shaded area obtained by the quadratic interpolation. Horizontal bars reflect the bin width of TKEL (25\,MeV) \cite{hildenbrand1983influence}.}
\end{figure}

\subsection{Fluctuations and correlation in TDRPA}

Next, we present the results of TDRPA calculations for the particle-number fluctuations and correlation. The particle-number fluctuations are quantified through Eq.~(\ref{eq:12}) for each impact parameter. Since the TKEL presented in Fig.~\ref{fig:1}(b) and $\sigma^2$ are calculated for each impact parameter, we can relate them to investigate the dissipation-fluctuation relationship. The results for the $^{144}$Sm+$^{144}$Sm and $^{154}$Sm+$^{154}$Sm reactions are presented in Figs.~\ref{fig:2}(a)(b) and \ref{fig:2}(c)(d), respectively. For comparison, TDHF-based fluctuations are computed with Eqs.~(\ref{eq:4}) and (\ref{eq:10}), which are shown in Fig.~\ref{fig:2}(a) and \ref{fig:2}(b). As mentioned above, the fluctuations in TDHF are significantly underestimated in the system $^{144}$Sm+$^{144}$Sm, both for the charge and mass numbers. In contrast, with increasing TKEL, the TDRPA fluctuations gradually deviate from those predicted by TDHF and closely reproduce the experimental values of $\sigma^2_{AA}$, see Fig.~\ref{fig:2}(a). It is important to notice, however, our calculations reveal that the charge fluctuation $\sigma_{ZZ}$ is overestimated by TDRPA, as shown in Fig.~\ref{fig:2}(b). The observed discrepancy may offer a room for refining terms in the EDF that are related to the mobility of protons (\textit{e.g.} shell effects or other time-odd contributions) with experimental data of low-energy heavy-ion reactions. Nonetheless, it is evident that TDRPA significantly improves the description of fluctuations and correlation within the microscopic mean-field approach.

\begin{figure}[t]
\includegraphics[width=0.42\textwidth]{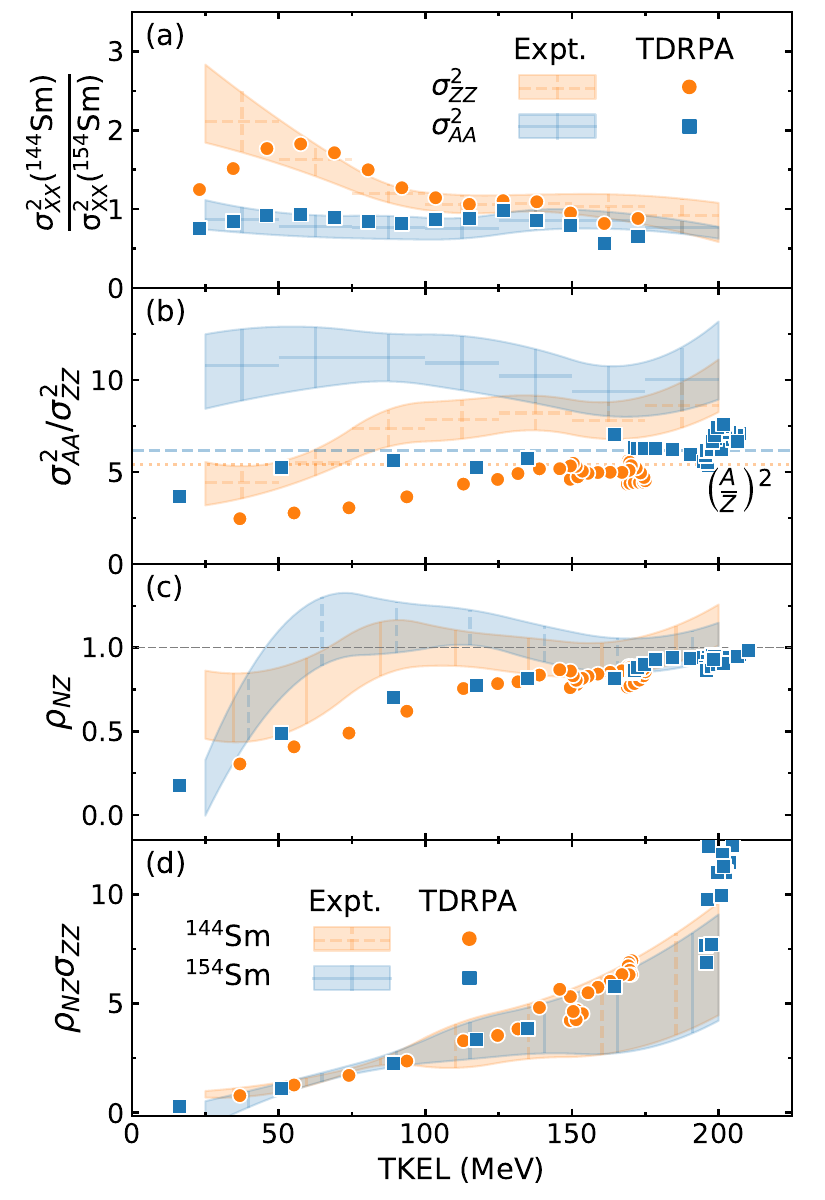}
\caption{\label{fig:3} In panel (a), the ratio of dispersions for $^{144}$Sm+$^{144}$Sm and $^{154}$Sm+$^{154}$Sm is shown as a function of TKEL. The theoretical (symbols) and experimental (lines with shaded area) ratios of $\sigma^2_{ZZ}(^{144}\text{Sm})/\sigma^2_{ZZ}(^{154}\text{Sm})$ and $\sigma^2_{AA}(^{144}\text{Sm})/\sigma^2_{AA}(^{154}\text{Sm})$ are shown in orange and blue colors, respectively. In panels (b), (c), and (d), the ratio of $\sigma^2_{AA}/\sigma^2_{ZZ}$, the correlation coefficient $\rho_{NZ}=\sigma_{NZ}^2/\sigma_{NN}\sigma_{ZZ}$, and $\rho_{NZ}\sigma_{ZZ}=\sigma_{NZ}^2/\sigma_{NN}$ are shown as functions of TKEL, respectively. Orange and blue colors are used for the $^{144}$Sm+$^{144}$Sm and $^{154}$Sm+$^{154}$Sm systems, respectively. Experimental data \cite{wu1981influence} are shown by lines with shaded area obtained by a quadratic interpolation. TDRPA results for the reaction of $^{144}$Sm+$^{144}$Sm and $^{154}$Sm+$^{154}$Sm are shown by circles and squares, respectively. In panel (b), the $(A/Z)^2$ value of the systems for $^{144}$Sm+$^{144}$Sm and $^{154}$Sm+$^{154}$Sm are indicated by orange-dotted and blue-dashed lines, respectively.}
\end{figure}

For deformed nuclei, relative orientations play an important role \cite{PhysRevC.101.034602}. However, such orientation effects can induce symmetry breaking in the colliding system, generating non-zero net nucleon transfer that fundamentally undermines the validity of TDRPA calculations \cite{PhysRevLett.54.1139}. In this study, both the Tip–Side and Side–Side orientations (the latter due to a small $\beta_3$ obtained from $^{154}$Sm) exhibit varying degrees of failure, resulting in significant unphysical fluctuations\footnote{We note that if one of $^{154}$Sm nuclei were rotated by $180^\circ$ on the reaction plane in our initial states, the symmetry should have been preserved even for the Side-Side configuration.}. In contrast, the Tip–Tip orientation yields results that are in good agreement with experimental observations, in which the symmetry is preserved for all times. Consequently, subsequent theoretical calculations will be principally focused on the Tip-Tip configuration to ensure physical reliability.

The distinct dynamical behaviors between the two nuclear systems are quantitatively characterized through comparative analysis of charge and mass distribution variances.  Figure~\ref{fig:3} shows the ratio of these variances, calculated by TDRPA for the two systems, alongside the experimental data. Here, a linear interpolation between TKEL and $\sigma^2$ is applied to obtain a uniform TKEL for evaluating the variance ratios in Fig.~\ref{fig:3}(a). Notably, the experimental ratio of $\sigma^2_{AA}$ exhibits weak TKEL dependence, maintaining a mean value of 0.82 that deviates slightly from $(144/154)^2 \approx 0.87$. This discrepancy implies an enhanced dynamical uncertainty in the deformed nuclear system, which can potentially be attributed to deformation-induced nucleon exchange complexity. Furthermore, in the low-TKEL region, TDRPA captures a pronounced enhancement of $\sigma^2_{ZZ}$ in the $^{144}$Sm+$^{144}$Sm system, especially for $\text{TKEL}\approx50$--$100$\,MeV. Several explanations for this phenomenon have been proposed in the Refs.~\cite{wu1981influence,hildenbrand1983influence}, with the consensus that the neutron shell closure in $^{144}$Sm hinders neutron transfer and thus favors proton transfer with low energy loss. The shell blocking mechanism gradually weakens with increasing energy dissipation, eventually disappearing. 

This also explains the behavior of $\sigma^2_{AA}/\sigma^2_{ZZ}$ between the systems as shown in Fig. \ref{fig:3}(b). However, the overestimation of $\sigma^2_{ZZ}$ by TDRPA, as displayed in Fig.~\ref{fig:2}, leads to a breakdown in the prediction of $\sigma^2_{AA}/\sigma^2_{ZZ}$ for individual systems, allowing only a qualitative description of the gradual reduction in differences between them. TDRPA yields $\sigma^2_{AA}/\sigma^2_{ZZ}$ values are lower than the experimental values, even if closer to the relation $\sigma^2_{AA}/\sigma^2_{ZZ} = (A/Z)^2$ at the highest TKEL given by both theoretical calculation \cite{beck1978mass} and some experimental results of other reaction systems \cite{rehm1981dissipative,schull1981influence,PhysRevLett.43.191,kratz1981primary}. The observed discrepancy requires further investigations, \textit{e.g.}, EDF dependence of the $\sigma_{AA}^2/\sigma_{ZZ}^2$ ratio, or processes that are missing in TDRPA such as two-body dissipations \cite{tohyama2016two,wen2018two}, which we leave for future works.

The systematic correlation between neutron and proton transfer dynamics is quantitatively examined in Fig.~\ref{fig:3}(c) through the normalized covariance coefficient, defined as $\rho_{NZ} = \sigma^2_{NZ}/\sigma_{NN}\sigma_{ZZ}$. Globally, the normalized correlation coefficient obtained from TDRPA show smaller values as compared with the experimental data, partly because of the overestimated $\sigma_{ZZ}$ values. One can see that the correlation between neutron and proton transfers increases with energy dissipation, reaching the full correlation (with $\rho_{NZ} \to 1$) at the highest TKEL. This behavior suggests a transition in the reaction mechanism from a quantum transport regime, where single-particle and shell structure play an important role, to a stochastic diffusion regime dominated by statistical nucleon exchanges with increasing energy dissipation.

To isolate the influence of overestimated $\sigma_{ZZ}$ in the correlation coefficient analysis, we introduce a new observable defined as $\rho_{NZ}\sigma_{ZZ} = \sigma_{NZ}^2/\sigma_{NN}$, with experimental verification presented in Fig.~\ref{fig:3}(d). The associated measurement uncertainties were systematically calculated through error propagation analysis. TDRPA calculations successfully reproduce the neutron-proton correlations observed in both the $^{144}$Sm+$^{144}$Sm system and the  $^{154}$Sm+$^{154}$Sm system at a wide TKEL region. At TKEL$\approx$ 200 MeV, the TDRPA maintains a systematic overestimation of experimental values, attributed to the combined effects of fully correlation ($\rho_{NZ} \to 1$) and amplified estimation of $\sigma_{ZZ}$.

\begin{figure}[t]
\includegraphics[width=0.42\textwidth]{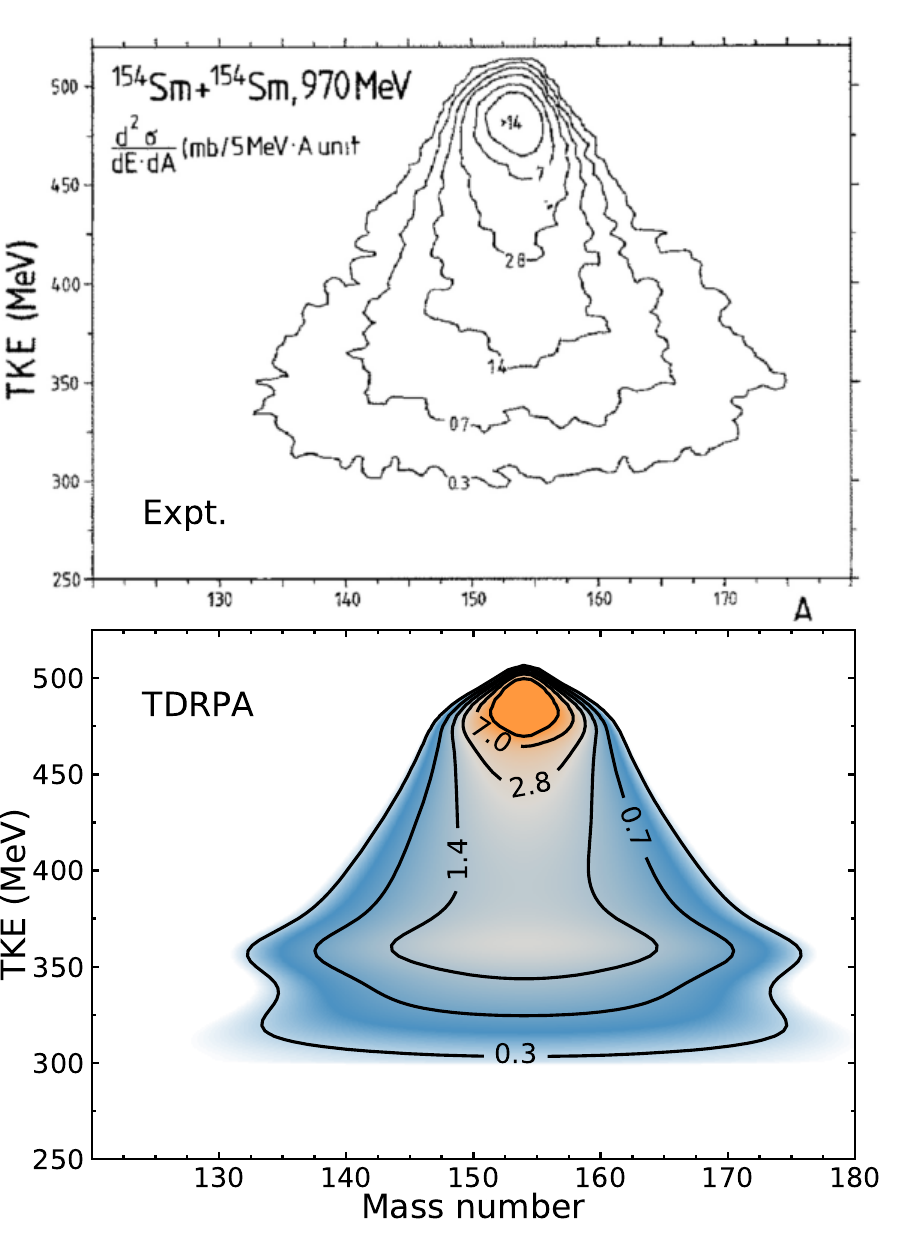}
\caption{\label{fig:4} Double-differential cross sections $\dd^2\sigma/\dd E\dd A$ for the $^{154}$Sm+$^{154}$Sm reaction at $E_{\rm c.m.} = 485$\,MeV are shown in the TKE-$A$ plane. Contour lines obtained from TDRPA calculations are shown in the lower panel, while experimental data taken from Ref.~\cite{hildenbrand1983influence} are shown in the upper panel.}
\end{figure}

\begin{figure*}[t]
\includegraphics[width=.75\textwidth]{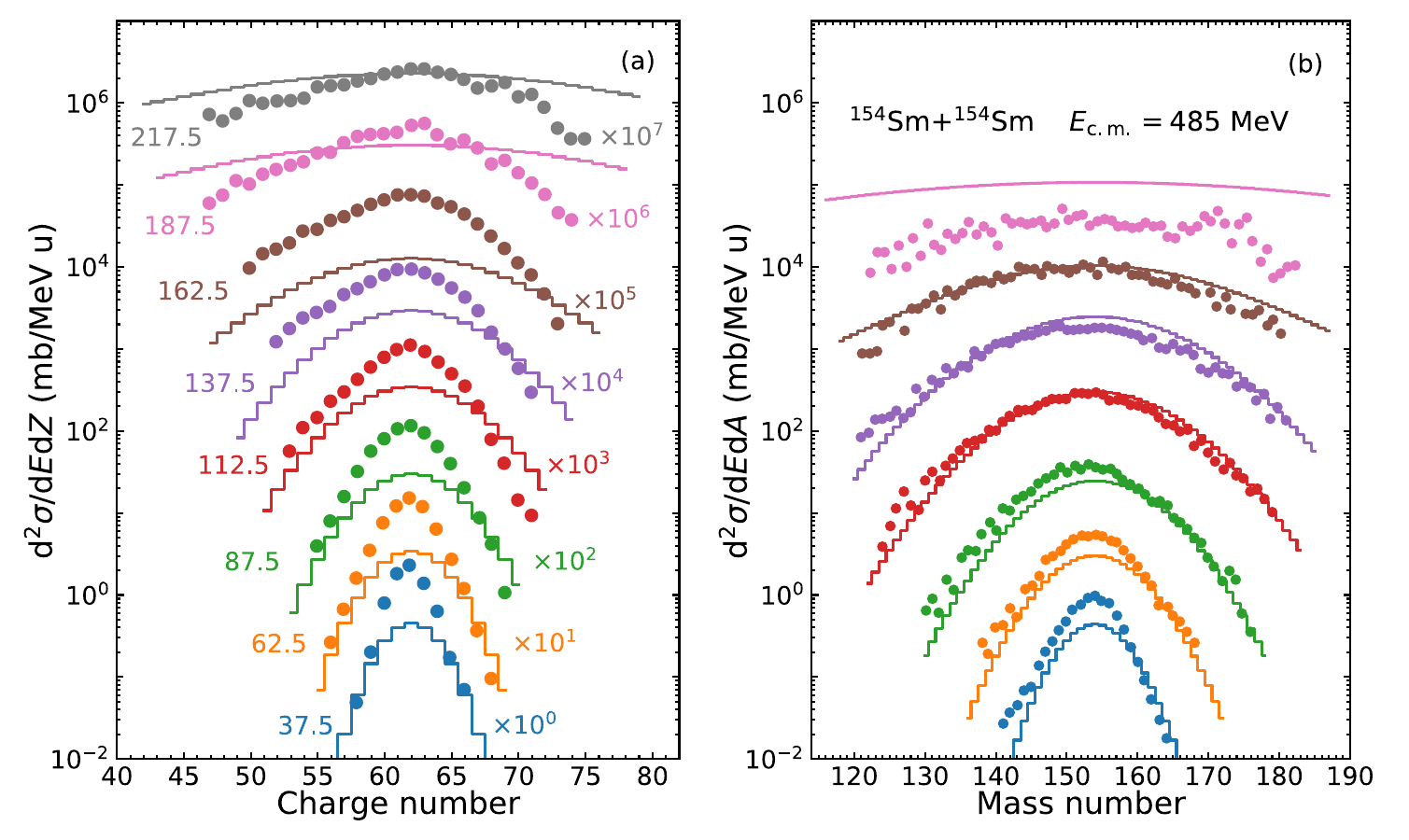}
\begin{minipage}{\textwidth}
\caption{\label{fig:5} Charge distributions (a) and mass distributions (b) of primary products for the $^{154}$Sm+$^{154}$Sm reaction at $E_{\rm c.m.} = 485$\,MeV are presented by different color for in each TKEL bin. The energy bins are 25-MeV wide. The midpoint value of each bin is indicated in the left panel on the left side of distributions. Both experimental (circles) \cite{hildenbrand1983influence} and theoretical (histograms) data are multiplied by the factors indicated in the left panel on the right side of distributions.}
\end{minipage}
\end{figure*}
\subsection{Mass and charge distribution in TDRPA}

We further investigate the distributions of reaction outcomes of the $^{154}$Sm+$^{154}$Sm reaction. Here, a Gaussian distribution is assumed to represent the probability distribution of the reaction products at each impact parameter. Given that TDRPA calculations did not account for momentum fluctuations, we assigned a TKE uncertainty of $\sigma_{E}=5$\,MeV, which corresponds to the experimental resolution \cite{hildenbrand1983influence}. Consequently, the bivariate Gaussian distribution in mass number ($A$) and TKE ($E$) for each impact parameter can be established as follows:
\begin{eqnarray}
\label{eq:15}
P(E,A) = \frac{1}{2\pi\sigma_{E}\sigma_{AA}} \exp\!\left[-\frac{(E-\mu_E)^2}{2\sigma_{E}^2} - \frac{(A-\mu_A)^2}{2\sigma^2_{AA}}\right],\nonumber\\[1mm]
\end{eqnarray}
where \(\mu_E\) and \(\mu_{A}\) are the corresponding mean values for the reaction. Note that $P(E,A)$ has the unit of MeV$^{-1}$, which represents a probability distribution with respect to $E$ and $A$ with the energy uncertainty of $\sigma_{E}=5$\,MeV. The double-differential cross section can subsequently be defined as:
\begin{equation}
\label{eq:16}
\frac{\dd^2\sigma}{\dd E\,\dd A}(E,A) = 2 \pi \int_{b_{\text{cut}}}^{b_{\text{max}}} P(E, A, b)\, b\, \dd b.
\end{equation}
The resulting double-differential cross sections are shown in Fig.~\ref{fig:4}. From the figure, we find that the TDRPA description of the double-differential cross sections is in excellent agreement with the experimental data. We emphasize that no normalization or adjustment was introduced to reproduce the experimental data, except the assumption of $\sigma_E$\,$=$\,5\,MeV TKE uncertainty. We note, however, that ideally the TKE fluctuation should also be determined based on microscopic calculations, such as the TDRPA \cite{PhysRevLett.54.1139} or SMF \cite{ayik2020kinetic} approach, which makes $\sigma_E$ impact-parameter dependent.
Note that only the mass distribution is compared, as the experimentally measured charge and neutron distributions are asymmetric, even though the detected fragments originate from a symmetric system. 

A more quantitative comparison of the double-differential cross sections is presented in Fig. \ref{fig:5}. Here, the theoretical results are obtained by integrating over consecutive TKEL intervals (each 25 MeV wide, ranging from 25 MeV to 225 MeV). Progressive broadening of mass and charge distributions with increasing energy loss reflects enhanced fluctuations associated with nucleon diffusion dynamics. The comparison of the mass distributions shown in Fig.~\ref{fig:5}(b) supports the validity of the Gaussian assumption. On the other hand, however, the experimental charge distribution deviates from a Gaussian shape as shown in Fig.~\ref{fig:5}(a), possibly due to interference from experimental background \cite{hildenbrand1983influence}. Overall, TDRPA demonstrates its capability to predict the cross sections of MNT products, thereby providing theoretical guidance for the synthesis of new isotopes in symmetric systems such as Yb+Yb \cite{PhysRevC.101.034602} and U+U \cite{ayik2017multinucleon}.

\section{Conclusions}
\label{sec:conclusion}

In this study, we implement the time-dependent Hartree-Fock (TDHF) and time-dependent random phase approximation (TDRPA) frameworks to systematically investigate deep-inelastic collisions of the symmetric $^{144}$Sm+$^{144}$Sm and $^{154}$Sm+$^{154}$Sm systems. The TDRPA method effectively addresses the inherent limitations of TDHF in capturing particle-number fluctuations and neutron-proton transfer correlations. While TDRPA demonstrates remarkable accuracy in reproducing experimental mass distributions of reaction products, it exhibits systematic overestimation of proton distribution widths, suggesting potential refinements in isospin-dependent interaction treatments. Notably, the TDRPA framework successfully resolves the characteristic suppression of nucleon transfer in closed-shell structure observed at low energy-loss regimes. This capability underscores its sensitivity to shell structure effects during the dissipation process. Our results establish TDRPA as a fully microscopic tool for probing fluctuation-dissipation mechanisms in heavy-ion collisions. Furthermore, the method provides critical predictive insights for MNT reaction, particularly for synthesizing neutron-rich exotic nuclei in symmetric systems.

\section{Acknowledgments}
This work was supported by the National Natural Science Foundation of China under Grants No. 12075327; The Open Project of Guangxi Key Laboratory of Nuclear Physics and Nuclear Technology under Grant No. NLK2022-01; Fundamental Research Funds for the Central Universities, Sun Yat-sen University under Grant No. 23lgbj003; International Program for Candidates, Sun Yat-sen University. The authors are grateful to the C3S2 computing center in Huzhou University for calculation support.


%

\end{document}